\documentclass[12pt,preprint]{aastex}
\usepackage{psfig}
 
\usepackage{graphicx}
\usepackage{epstopdf}
\usepackage{psfig}

\newcommand{\etal}{\mbox{et~al.}}

\newcommand{\msun}{\mbox{$M_{\odot}$~}}

\newcommand{\lsun}{\mbox{$L_{\odot}$~}}

\def\deg      {{\ifmmode^\circ\else$^\circ$\fi}} 

\def\kms   {{\ km\ s$^{-1}$} }
 
\slugcomment{To appear in the Astrophysical Journal.}
 
\shorttitle{Imaging the molecular gas in a submm galaxy at $z = 4.05$}
\shortauthors{Carilli et al.}
 
\begin{document}
  
 \title{Imaging the molecular gas in a submm galaxy at $z = 4.05$:
cold mode accretion or a major merger?}
 
\author{ 
C.L. Carilli\altaffilmark{1},
E. Daddi\altaffilmark{2},
D. Riechers\altaffilmark{3},
F. Walter\altaffilmark{4},
A. Weiss\altaffilmark{5},
H. Dannerbauer\altaffilmark{2},
G.E. Morrison\altaffilmark{6},
J. Wagg\altaffilmark{7},
Romeel Dav\'e\altaffilmark{8},
D. Elbaz\altaffilmark{2},
D. Stern\altaffilmark{9},
M. Dickinson\altaffilmark{10},
M. Krips\altaffilmark{11},
M. Aravena\altaffilmark{1}
}

\altaffiltext{$\star$}{The Very Large Array of the National Radio Astronomy
Observatory, is a facility of the National Science Foundation
operated under cooperative agreement by Associated Universities, Inc}

\altaffiltext{1}{National Radio Astronomy Observatory, P.O. Box 0, 
Socorro, NM, USA 87801-0387}
\altaffiltext{2}{Laboratoire AIM, CEA/DSM - CNRS - University Paris Diderot, 
DAPNIA/Service Astrophysccique, CEA Saclay, Orme des Merisiers, 
91191 Gif-sur-Yvette, France}
\altaffiltext{3}{Department of Astronomy, California Institute of Technology,
MC 249-17, 1200 East California Boulevard, Pasadena, CA 91125, USA; Hubble 
Fellow}
\altaffiltext{4}{Max-Planck Institute for Astronomy, Konigstuhl 17, 69117,
Heidelberg, Germany}
\altaffiltext{5}{Max-Planck Institute for Radio Astrnomy, 
Auf dem Hügel 69, 53121 Bonn, Germany}
\altaffiltext{6}{Institute for Astronomy, University of Hawaii, Honolulu, HI,
96822, USA ; Canada-France-Hawaii Telescope, Kamuela, HI 96743}
\altaffiltext{7}{ESO-ALMA, Alonso de Cordova 3107 Vitacura Casilla
19001 Santiago 19, Chile}
\altaffiltext{8}{Steward Observatory, University of Arizona, 
Tucson, AZ, USA 85721}
\altaffiltext{9}{Jet Propulsion Laboratory, California Institute of 
Technology, Pasadena, CA 91109, USA}
\altaffiltext{10}{National Optical Astronomy Observatory, 
950 North Cherry Ave,  Tucson, AZ, USA, 85719}
\altaffiltext{11}{IRAM, Domaine Universitaire, 38406 St-Martin-d'Hères, 
France}
 
\begin{abstract}

We present a high resolution (down to $0.18"$), multi-transition
imaging study of the molecular gas in the $z = 4.05$ submillimeter
galaxy GN20.  GN20 is one of the most luminous starburst galaxy known
at $z > 4$, and is a member of a rich proto-cluster of galaxies at $z
= 4.05$ in GOODS-North.  We have observed the CO 1-0 and 2-1 emission
with the VLA, the CO 6-5 emission with the PdBI Interferometer, and
the 5-4 emission with CARMA. The H$_2$ mass derived from the CO 1-0
emission is $1.3 \times 10^{11} (\alpha/0.8)$ M$_\odot$. High
resolution imaging of CO 2-1 shows emission distributed over a large
area, appearing as partial ring, or disk, of $\sim 10$kpc diameter.
The integrated CO excitation is higher than found in the inner disk of
the Milky Way, but lower than that seen in high redshift quasar host
galaxies and low redshift starburst nuclei. The CO 4-3 integrated line
strength is more than a factor of two lower than expected for thermal
excitation. The excitation can be modeled with two gas components: a
diffuse, lower excitation component with a radius $\sim 4.5$kpc and a
filling factor $\sim 0.5$, and a more compact, higher excitation
component (radius $\sim 2.5$kpc, filling factor $\sim 0.13$). The
lower excitation component contains at least half the molecular gas
mass of the system, depending on the relative conversion factor.  The
VLA CO 2-1 image at $0.2"$ resolution shows resolved, clumpy
structure, with a few brighter clumps with intrinsic sizes $\sim 2$
kpc. The velocity field determined from the CO 6-5 emission is
consistent with a rotating disk with a rotation velocity of $\sim 570$
km s$^{-1}$ (using an inclination angle of 45$^o$), from which we derive
a dynamical mass of $3 \times 10^{11}$ \msun within about 4 kpc
radius. The star formation distribution, as derived from imaging of
the radio synchrotron and dust continuum, is on a similar scale as the
molecular gas distribution.  The molecular gas and star formation are
offset by $\sim 1"$ from the HST I-band emission, implying that the
regions of most intense star formation are highly dust-obscured on a
scale of $\sim 10$kpc. The large spatial extent and
ordered rotation of this object suggests that this is not a major
merger, but rather a clumpy disk accreting gas rapidly in minor
mergers or smoothly from the proto-intracluster medium.
Qualitatively, the kinematic and structural properties of GN20 compare
well to the most rapid star-formers fed primarily by cold accretion in
cosmological hydrodynamic simulations. Conversely, if GN20 is a major,
gas rich merger, then some process has managed to ensure that the star
formation and molecular gas distribution has not been focused into
one or two compact regions.

\end{abstract}
 
  
\section{Introduction}

Studies of the stellar populations of elliptical galaxies imply that
massive ellipticals form the bulk of their stars fairly quickly
(timescales $\le 1$ Gyr) at early epochs ($z > 2$).  Moreover, there
is a clear trend with increasing mass such that the more massive the
galaxy, the earlier and quicker the star formation (see review by
Renzini 2006).  This conclusion is supported by studies of specific
star formation rates (SFR/stellar mass), indicating `downsizing' in
galaxy formation, with active star formation being preferentially
quenched in more massive galaxies over cosmic time (Noeske et al.
2007, 2009; Zheng et al. 2007; Pannella et al. 2009), as well as the
direct observation of old stellar populations in early type galaxies
at $z \ge 1$, implying formation redshifts $z > 3$ (Collins et
al. 2009; Kurk et al. 2009, Kotilainen et al. 2009; Papovich et al.
2010). These results imply that there should be a progenitor
population of active, clustered star forming galaxies at high
redshift.

Bright submm-selected galaxies (SMGs; $S_{850\mu m} > 5$ mJy; see
Blain et al. 2002 for a review) are an important class of source in
this regard. Although they are relatively rare, with typical space
densities of $10^{-5}$--$10^{-6}$~Mpc$^{-3}$, they have very high
bolometric luminosities ($\sim 10^{13}$ \lsun), implying the most
intense bursts of star formation known ($\sim 1000$ \msun yr$^{-1}$).
While radio galaxies and quasars can reach similar or higher
luminosities (Miley \& de Breuck 2007; Solomon \& vanden Bout 2006),
the latter objects contain bright active galactic nuclei (AGN), and it
is possible that some of their far-infrared emission is powered by
accretion onto black holes. SMGs, on the other hand, are known to be
largely star formation dominated galaxies (Alexander et al. 2005).

The emerging scenario is that SMGs may be the starburst progenitors of
massive early type galaxies.  These hyper-luminous high-z galaxies
often trace high overdensities (Stevens et al.  2003; Aravena et
al. 2009; although cf. Chapman et al.  2009), and are likely related
to the formation of giant elliptical galaxies in clusters. A key
question for the SMGs is: what drives the prolific star formation?
Tacconi et al. (2006; 2008) argue, based on CO imaging of a sample of
$z \sim 2$ SMGs, that SMGs are predominantly nuclear starbursts, with
median sizes $< 0.5"$ ($< 4$kpc), 'representing extreme, short-lived,
maximum star forming events in highly dissipative mergers of gas rich
galaxies.'  This conclusion is supported by VLBI imaging of the star
forming regions in two SMGs (Momjian et al. 2005; 2009; 2010).  We
return to this question below.

The discovery of apparently old elliptical galaxies at $z \ge 2$ has
pushed the question of starburst progenitors of giant elliptical
galaxies to even earlier epochs (Cimatti et al 2004; Wikind et al.
2008; Mobasher et al. 2005; Kriek et al. 2008; Doherty et
al. 2009). The redshift distribution for about 50\% of the SMG
population, namely, the radio detected sources, has been shown to peak
around $z \sim 2.3$, with most of these sources being between $z \sim
1.5$ and 3 (Carilli \& Yun 2000; Chapman et al. 2003; Wagg et
al. 2009). However, there is a low redshift bias in radio-selected
samples, and the question remains: is there  a  substantial
($\sim 30\%$) population of SMGs at $z > 3$?  These higher redshift
sources would potentially pin-point very early formation of the most
massive ellipticals.  Early searches for SMGs at $z>4$ (Dannerbauer et
al. 2002; 2004; 2008; Dunlop et al. 2004; Younger et al. 2007; 2008;
Wang et al. 2007), were unsuccessful. However, recently, a number of
SMGs have been found at $z > 4$, including two in the COSMOS field ($z
= 4.5$ and 4.7; Capak et al. 2008; Schinnerer et al. 2008, 2009), GN10
at $z \sim 4.04$ in GOODS-North (Daddi et al. 2009b), a $z = 4.76$ SMG
in the CDF-South (Coppin et al. 2009), a strongly lensed source at $z
= 4.044$ (Knudsen et al. 2009), and GN20, the subject of this
paper. Daddi et al. (2009a) conclude, based on SMG space densities and
duty cycles, that there are likely enough SMGs at $z > 3.5$ to account
for the known populations of old massive galaxies at $z \sim 2$ to
3. They also point out that the contribution of SMGs to the comoving
cosmic star formation rate density at $z \sim 4$ (SFRD $\sim 0.02$
M$_\odot$ yr$^{}$ Mpc$^{-3}$) is comparable to that of Lyman break
galaxies.

\section{The case of GN20: an ideal laboratory for studying clustered
galaxy formation within 1.5 Gyr of the Big Bang}

In this paper, we present the most detailed, multi-transition, high
resolution imaging study of CO emission from an SMG to date. Our study
focuses on a recently discovered SMG in GOODS North, GN20, at $z =
4.05$.  GN20 was originally detected at 850$\mu$m by Pope et al.
(2006) with a flux density of $S_{850\mu m}=20.3$mJy. The SED from
optical through radio wavelengths is well sampled, and a detailed
analysis implies a hyper-luminous infrared galaxy with a total IR
luminosity (8 to 1000$\mu$m) of $L_{IR} = 2.9\times 10^{13}$
L$_\odot$, and a dust temperature $\sim 57$K (Daddi et al. 2009a). The
SED is consistent with a star forming galaxy, with a total star
formation rate (0.1 to 100 M$_\odot$) of 3000 M$_\odot$ yr$^{-1}$ (see
Section 5.1).

The HST I-band image of GN20 shows diffuse emission about 1.5$"$ in
length (Daddi et al. 2009a), although offset from the radio and submm
emission by $\sim 1"$.  Daddi et al. (2009a) derive a stellar mass is
$2.3\times 10^{11}$ \msun from IR through optical SED fitting. The
rest-frame UV spectrum shows no lines, typical for SMGs, but detailed
study of the broadband SEDs are consistent with the observed CO
redshift (Daddi et al.  2009a).

GN20 is detected in 1.4GHz continuum emission (Morrison et al. 2009)
using the VLA, and a combination of the VLA + MERLIN at higher
resolution (Casey et al. 2009). The source is resolved on a scale of
$\sim 1.5"$, with a total flux density at 1.4GHz of $72\pm13\mu$Jy. We
re-analyze the 1.4GHz data in Section 4.4 below.  High resolution
imaging of the 850$\mu$m emission also shows resolved structure, with
a north-south extension possibly as large at 1.5$"$ (Younger et al.
2008; Iono et al. 2006). Again, we return to these data below.  The
6.2$\mu$m PAH spectral feature has been detected in GN20 using
Spitzer (Riechers et al. in prep).

GN20 resides in a dense cosmic environment. Most prominent are a pair
of SMGs located about 20$"$ to the southwest of GN20, designated
GN20.2a and GN20.2b. The total 850$\mu$m flux density for these two
sources is 9.9 mJy.  Daddi et al. (2009a) also find 15 B-band dropout
galaxies in a 25$''$ radius centered on GN20, an overdensity of a
factor of 6 compared to the full GOODS-N area, which is significant at
the 7$\sigma$ level.  A spike in the redshift distribution of galaxies
at $z=4.06\pm0.02$ is observed in all of GOODS-N (13 spectroscopic
redshifts in total at this redshift).  Lastly, the SMG GN10 at $z =
4.04$ is located about 9$'$ from GN20 (Daddi et al. 2009b).
Therefore, it appears that the GN20 volume has a very significant
overdensity, indicating a proto-cluster environment at
$z\sim 4.05$. Daddi et al. (2009a) estimate a total mass for this
structure of $\sim 10^{14}$ \msun. This proto-cluster presents an
ideal opportunity to study massive galaxy formation within $\sim
1.5$Gyr of the Big Bang.
 
GN20 has been detected in CO 4-3 emission, indicating large
amounts of molecular gas, the requisite fuel for star formation (Daddi
et al. 2009a).  GN20 was not detected in CO 7-6 or [CI] 809GHz
emission (Casey et al. 2009).

In this paper we present high resolution imaging of the CO emission
from GN20. We observed the low order transitions (1-0 and 2-1) with
the Very Large Array (VLA) down to 0.18$"$ resolution.  The low order
transitions are critical for determining the total gas mass, since low
redshift conversion factors of CO luminosity to H$_2$ mass are
calibrated using the low order transitions. We observed the higher
order transitions (CO 6-5 and 5-4) with the Plateau de Bure
Interferometer (PdBI) and the Combined Array for Research in
Millimeter Astronomy (CARMA).  We consider the gas excitation and
physical conditions in the molecular gas, and we compare these to low
and high redshift galaxies.
 
\section{Observations} 

\subsection{Very Large Array} 

We observed the GN20 field with the Very Large Array in the B (10km),
C (3km), and D (1km) configurations. Observations were made of both
the CO 2-1 line and the CO 1-0 line.

Figure 1 shows the spectral coverage of the 2-1 and 1-0 VLA
observations relative to the CO 4-3 line profile.  The CO 2-1
observations were done using two channels of 50 MHz bandwidth each,
and two polarizations each. The two channels were centered at 45.585
and 45.635 GHz. These were the optimal IF settings to cover $\sim 80\%$
of the CO line profile, as determined from the 4-3 or 6-5, given the
current correlator limitations at the VLA. These IFs miss the two
edges of the emission line with this tuning.  A total of about 70
hours were spent on-source for the 2-1 observations over the three
configurations.  About 15\% of the time was spent observing the
continuum at 43GHz using 100 MHz bandwidth.

The CO 1-0 emission was observed using two channels of 50 MHz
bandwidth each, and two polarizations each, with the channels centered
on 22.815 GHz and 22.935 GHz in the D array. The total observing time
was about 20 hour.  The first frequency setting covers most of the
emission line ($\sim 75\%$, missing the low frequency 
edge of the line), as shown in Figure 1.  The second setting was
used to obtain a sensitive limit to the continuum.

Fast switching phase calibration was employed (Carilli \& Holdaway
1999) on timescales between two and three minutes using the VLA
calibrator J1302+5748.  Data were edited to remove time ranges of poor
phase stability.  The source 3C286 was used for flux density
calibration. The full resolution of the B array at 23 and 45 GHz is
0.4$"$ and 0.2$"$, respectively. By including C and D array data, we
have good UV coverage for structures up to scales of 16$''$ and 8$''$,
respectively.  After calibration and data editing, we synthesized
images at various resolutions to investigate structure over this range
in spatial scale. During the image processing, the synthesized beam
was deconvolved down to CLEAN residuals $\sim 1\sigma$ in a CLEAN box
centered on the galaxy.

\subsection{Plateau de Bure Interferometer}

We observed the CO 6-5 line toward GN20 in 2009 January and 2008 May,
using the PdBI in its B and D configurations.  At $z$=4.055, this line
is redshifted to 136.7899\,GHz. Weather conditions were good for 2\,mm
observations. The calibration is estimated to be good within 15\%.

The 2\,mm receivers were tuned to 136.97\,GHz to cover both the
emission from GN20 and a lower-$J$ CO line from a nearby $z$=1.5
galaxy. D configuration observations were pointed close to this nearby
galaxy, resulting in a primary beam attenuation (PBA) of 2.24 at the
position of GN20. B configuration observations were pointed in between
GN20 and its companions, GN20.2a and GN20.2b, resulting in a PBA of
1.076 at the position of GN20. Observations were carried out in dual
polarization mode, covering a total bandwidth of 1\,GHz.  For
calibration of the data we observed standard bandpass calibrators
(J0418+380, 3C273), phase/amplitude calibrators (J1044+719,
J1150+497), and flux calibrators (MWC 349, 3C273).

For data reduction and analysis, the GILDAS package was used. To
extract the spectral data, a combined B and D configuration dataset
was used to optimize sensitivity. The spectrum was extracted directly
from the uv data, fitting a circular Gaussian to the
velocity-integrated line emission and then extracting the spectral
data over the same area. For imaging, only B configuration data was
used to optimize beam shape and resolution. After editing, the uv data
were imaged with both natural (NA) and uniform (UN) baseline
weightings, leading to resolutions of 0.89$''$$\times$0.76$''$ and
0.84$''$$\times$0.67$''$. The CLEAN algorithm was used for
deconvolution, and applied down to 1.5$\sigma$ in a small box centered
on the galaxy. The sensitivity achieved with the B\&D configuration
dataset is 0.41\,mJy\,beam$^{-1}$ per 25\,\kms\ channel (NA). The
sensitivity of the B configuration dataset is 0.22\,mJy\,beam$^{-1}$
per 150\,\kms\ channel (UN), and 0.081/0.096\,mJy\,beam$^{-1}$ per
800\,\kms\ channel (NA/UN).

\subsection{CARMA}

We observed the CO 5-4 transition line toward GN20 on 2009 July 10,
using CARMA in E array. At $z$=4.055, this line is redshifted to
113.999\,GHz. Weather conditions were good for 3\,mm observing.  The
nearby source 0958+655 was observed every 15\,minutes for secondary
amplitude and phase calibration. 3C273 was observed for bandpass
calibration, and fluxes were bootstrapped relative to MWC349. Pointing
was performed every 2\,hr on nearby sources, using both radio and
optical modes. The resulting calibration is estimated to be accurate
within 15--20
redshifted line frequency in the upper sideband (USB). The widest
correlator mode was used, overlapping the three bands with
15\,channels of 31.25\,MHz width by 2 edge channels each to improve
relative calibration. This leads to an effective bandwidth of
1281.25\,MHz per sideband. For data reduction and analysis, the MIRIAD
package was used.

\section{Results} 

\subsection{CO 1-0}

Figure 2 shows the image of the CO 1-0 emission made using the D array
data at $3.7"$ resolution.  The crosses in this image show the optical
positions of GN20, GN20.2a, GN20.2b, plus a $z = 4.05$ Lyman-break
galaxy J123711.53+622155.7.  Results from all of our CO observations
are summarized in Table 1.

GN20 is detected in CO 1-0 emission, with a peak surface brightness of
$0.19\pm 0.03$ mJy beam$^{-1}$ at J123711.95+622212.2.  The error
quoted is the rms noise on the image (24$\mu$Jy) plus a $\sim 10\%$
uncertainty in flux calibration estimated from the variation of the
bootstrapped flux density of the phase calibrator on subsequent days,
added in quadrature.  A Gaussian fit to GN20 implies a slightly
resolved source, with a total flux density of $0.25\pm 0.05$ mJy, and
a deconvolved source size FWHM = $2.8" \times 1.4"$, with a major axis
position angle north-south. There is an interesting diffuse structure
extending southwest of GN20, and north of both GN20.2a and b, with a
mean surface brightness of about 60 $\mu$Jy beam$^{-1}$, or about
$2\sigma$, over roughly 10$"$. Deeper observations in D array are
required to confirm the reality of this diffuse component.

The total velocity range covered is 670 km s$^{-1}$, and the implied
velocity integrated line intensity is $0.22\pm0.04$ Jy km s$^{-1}$, after
correcting for the 25\% of the line that falls outside the band.

No continuum emission is seen from GN20 in the off-line 23GHz 
image to a 1$\sigma$ limit of 30$\mu$Jy.

\subsection{CO 2-1}

We have synthesized images of CO 2-1 emission from GN20 at three
different spatial resolutions in order to investigate structure on
scales ranging from $0.18"$ up to a few arcsec. Figure 3 shows images
made from the B+C+D data using a Gaussian taper of the UV data of 400
k$\lambda$, 900 k$\lambda$, and no taper, from top to bottom. Natural
weighting (Cornwell, Braun, \& Briggs 1999) was used in all cases. The
corresponding resolutions are $0.45''$, $0.25''$, and $0.18''$,
respectively.

At $0.45"$ resolution (Figure 3a) the source is detected, with a total
flux density in the CO 2-1 image of 0.73 mJy. The rms in this image is
36$\mu$Jy beam$^{-1}$. The implied velocity integrated line intensity
is $0.74\pm 0.1$ Jy km s$^{-1}$, after correcting for the 20\% of the
line that falls outside the band. The source is well resolved,
appearing as an incomplete ring or disk-like structure with a diameter
of $\sim 1.5"$. 

The higher resolution images (Figures 3b, c) show well resolved
substructure. A Gaussian fit to the most compact knot in the $0.25"$
resolution image, on the southern part of the disk, yields a
deconvolved size of $0.27" \times 0.15"$, a peak surface brightness of
0.14 mJy beam$^{-1}$, and a total flux density of 0.24 mJy $\sim 30\%$
of the total CO 2-1 emission. The implied rest frame brightness
temperature of the clump is 12K. Resolving the structure on these
scales has the important implication that the CO emission in GN20 is
not dominated by one or two compact ($<$ 1kpc), high brightness
temperature regions, but the emission must come from numerous, smaller
clouds distributed over the disk. For comparison, recent imaging of
the CO2-1 emission at similar resolution in the host galaxy of the $z
= 4.4$ quasar BRI 1335-0417 yielded a maximum intrinsic brightness
temperature of 60K (Riechers et al. 2008).

No continuum emission is seen from GN20 in the 43GHz image at 
$1.75"$ resolution to a 1$\sigma$ limit of 60$\mu$Jy beam$^{-1}$. 

\subsection{CO 6-5 and 5-4}

The Plateau de Bure Interferometer integrated spectrum of the CO 6-5
emission from GN20 is shown in Figure 1. A Gaussian model fit to the
spectral profile results in a FWHM $= 664 \pm 50$ km s$^{-1}$, and an
integrated flux of $1.8 \pm 0.2$ Jy km s$^{-1}$. However, the profile
itself looks more flat-topped than a Gaussian.

Figure 4 shows the channel images at 150 km s$^{-1}$ spectral
resolution, and 0.8$"$ spatial resolution (UN weighting). Figure 5a
shows the total CO 6-5 emission integrated over 800 km s$^{-1}$ (NA
weighting), while Figure 5b shows the moment 1 image (the
intensity-weighted velocity centroid; UN weighting).  The emission is
resolved in space and velocity.  A Gaussian fit to the total emission
image results in a (deconvolved) source major axis of FWHM $= 0.72"
\pm 0.06"$.  The velocity channels show clear motion from northeast to
southwest, with the centroid of the emission moving by about $1.0"\pm
0.2"$ over the 750 km s$^{-1}$ range.

We have convolved the CO 2-1 emission to the resolution of the 6-5
images. A Gaussian fit to the convolved 2-1 image results in an
intrinsic source FWHM $1.2"\pm 0.15''$, as expected given the disk
seen at high resolution (Figure 3). Hence, at matched resolution, the
Gaussian fit CO 2-1 source size is about a factor 1.7 larger than CO
6-5. While both the low and high order CO emission is spatially
extended, there is evidence that the high order emitting regions are
smaller than the low order emitting regions.

We have marginally detected the CO 5-4 emission from GN20 using
CARMA. The signal-to-noise is only moderate, but we are able to derive
a velocity integrated flux of $2.2 \pm 0.7$ Jy km s$^{-1}$ by
averaging the channels defined by the 6-5 line profile.

The integrated continuum emission from GN20 at 137 GHz is $0.89 \pm
0.15$ mJy. The continuum source is resolved, with a Gaussian size
of FWHM $= 0.72" \pm 0.15"$.

\subsection{Comparison to optical and radio imaging}

We compare the high resolution CO imaging herein with observations at
other wavelengths, and compare the CO 6-5 with the CO 2-1 images.

Figure 6a shows an overlay of the CO 2-1, the CO 6-5 emission, and the
HST i-band image (Giavalisco et al. 2005). The CO 2-1 emitting region
has a major axis $\sim 1.5"$ area.  The optical emission is complex,
with a North-South filament centered about $1"$ west of the CO
emission and extending along the western edge of the CO emission for
about $\sim 1.5"$. Generally, the CO falls in a minimum in the HST
image, implying substantial obscuration over a scale $\sim
10$kpc. Note that GOODS astrometry relative to the radio imaging has
been shown to be accurate to better than $0.15"$ (Daddi et al. 2007).

Figure 6b shows the same CO 2-1 and HST images, but with CO 6-5
replaced by the SMA 890$\mu$m image of the rest-frame FIR continuum
(Younger et al. 2008).  Again, the FIR continuum emission comes from
the CO emitting regions, well-offset from the rest-frame UV emission
seen by the HST.  The SMA image is lower resolution, but the dust
emission appears to be extended in a similar manner to the CO, and
fitting to the visibilities by Younger et al.  (2008) implies a source
that is at least $0.6"$ in intrinsic extent, and possibly as large as
$1.5"$.

Figure 6c shows the HST I-band and CO 2-1 images along with the IRAC
3.6$\mu$m image (sampling the rest frame I-band). The IRAC image has a
PSF with FWHM $=1.5"$. The IRAC peak is offset by $\sim 1"$ from the
ACS imaging, and coincident with the CO peak.

We have reanalyzed the VLA, and VLA+MERLIN, 1.4 GHz images of GN20
(Morrison et al. 2009; Casey et al. 2009).  Figure 7 compares the CO
2-1 emission and the 1.4 GHz emission from the combined VLA+MERLIN
data, both convolved to 0.45$"$ resolution.  A Gaussian fit to the VLA
image at $1.7"$ resolution implies a resolved source with a peak
surface brightness of $43 \pm 5\mu$Jy beam$^{-1}$ at
J123711.885+622211.79 and a total flux density of $72\pm13\mu$Jy. The
deconvolved source size is roughly circular, with a FWHM $\sim 1.4"$.

The combined MERLIN+VLA image has a resolution of FWHM = $0.41"$
(Casey et al.  2009). Most of the emission is resolved-out at this
high resolution.  A Gaussian fit to the brightest knot in the image to
the north yields a peak surface brightness of $14 \pm 4\mu$Jy at
J123711.941+622212.43, and a total flux density of $26\pm 9\mu$Jy.
While relatively low signal to noise, the 1.4 GHz emission, presumably
tracing star formation, shows a similar distribution to the CO 2-1
emission. The brightest radio peak is in the north, close to the CO
knot, and showing a similar extension east-west. A second 1.4 GHz peak
is seen about 1$"$ to the south, roughly coinciding with southern CO
peak. We emphasize that these two 'peaks' comprise only about 1/3 of
the total radio emission, implying that the 1.4GHz emission must be
diffuse, and distributed over the disk on a scale $\ge 1"$.

\section{Analysis}

The observed and derived parameters for GN20 are listed in Table 1.
We discuss these derivations, and other issues, in the sections below.

\subsection{Star formation rate and surface density} 

Daddi et al. (2009a) derive a total IR luminosity from detailed SED
fitting of $L_{IR} = 2.9\times 10^{13}$ L$_\odot$.
The implied total star formation rate is 3000 M$_\odot$ yr$^{-1}$,
using a Chabrier (2003) IMF.\footnote{Using a Salpeter IMF from 0.1 to
100 M$_\odot$, which is less top-heavy than the Chabrier IMF, would
increase this star formation rate by a factor $\sim 1.7$.}  This is
also consistent with the radio flux density, assuming the standard
relationship between radio luminosity and star formation rate (Yun et
al. 2001).

Normalizing by the stellar mass gives a specific star formation rate
of $1.3 \times 10^{-8}$ yr$^{-1}$, comparable to $z \sim 2$ SMGs,
but an order of magnitude higher than LBGs at $z \sim 4$ (Daddi et
al. 2009a). The stellar mass in GN20 was derived by Daddi et al (2009a)
based on the broad-band SED, including the IRAC photometry, which
samples the rest-frame I-band, plus the shorter-wavelength images. In
Figure 6c we found that the 3.6$\mu$m emission was co-spatial with the
CO. However, in Figure 6a and b we found the HST-ACS emission was
clearly offset from the CO, radio continuum, and dust, ie. from the
regions of highest star formation rate.  Hence, obscuration clearly
affects the SED, in particular at observed optical wavelengths, and
caution should be used when deriving the the stellar mass from
integrated SED fitting.

Imaging of the molecular gas, radio continuum, and rest-frame thermal
dust continuum emission in GN20 are all consistent with the gas and
star formation being distributed over a disk between 2.5 kpc and 4.5
kpc in radius.  The average star formation rate per unit area is then
roughly 60 \msun yr$^{-1}$ kpc$^{-2}$, with a peak a factor three
higher.  This value is well below the Eddington-limited "maximal
starburst rate" of 1000 \msun yr$^{-1}$ kpc$^{-2}$, ie. self-limited
star formation due to radiation pressure on the dust grains in a
self-gravitating gas disk (Thompson 2008). Such a maximal condition
has been seen in the star forming cores of Galactic Giant Molecular
Clouds (GMCs), the nuclei of nearby Ultraluminous Infrared
Galaxies,\footnote{ULIRGs are defined as galaxies with $L_{FIR} \sim
10^{12}$ L$_\odot$. These galaxies are typically luminous nuclear
starbursts (Downes \& Solomon 1998)} and in high redshift quasar host
galaxies (Walter et al. 2009).  For comparison, Tacconi et al. (2006)
derived a typical value of 120 \msun yr$^{-1}$ kpc$^{-2}$ for $z \sim
2$ SMGs (corrected to the Chabrier IMF).  Overall, it appears that
GN20 is forming stars at a rate comfortably below the value necessary
to disrupt the disk due to radiation pressure on dust.

\subsection{Gas and dust mass, and surface density}

Low order CO transitions are required to obtain an accurate estimate
of the total gas mass, since the low redshift conversion factor of CO
to H$_2$ mass is based on CO 1-0 emission (Downes \& Solomon 1998).
We find the velocity integrated CO luminosity for GN20 is: L$'_{\rm CO
1-0} = 1.6 \times 10^{11}$ K \kms pc$^{2}$, or $7.8\times 10^6$ \lsun.

Converting L$'_{\rm CO 1-0}$ into an H$_2$ mass requires a mass
conversion factor, $\alpha$.  Unfortunately, this conversion factor
varies across different galaxy types, and possibly within a given
galaxy.  Values for $\alpha$ range from $\sim 3.4$ M$_\odot$ (K km
s$^{-1}$ pc${^2}$)$^{-1}$ for Milky Way GMCs, to 0.8 M$_\odot$ (K km
s$^{-1}$ pc${^2}$)$^{-1}$ for nearby ULIRGs (Downes \& Solomon 1998;
Solomon \& Vanden Bout 2006; Daddi et al. 2010a). The value is likely
a function of metalicity, excitation, and ISM pressure (Leroy et al.
2008; Bigiel et al. 2008). For SMGs, previous studies have uniformly
assumed the low redshift ULIRG value, in which case we obtain an H$_2$
mass of $1.3\times 10^{11} \times (\alpha/0.8)\msun$ for GN20.  The
gas mass is 57\% of the stellar mass estimated by Daddi et
al. (2009a), although again, caution is needed when deriving stellar
masses in the presence of substantial obscuration.

Our gas masses are higher than those derived by Daddi et al. (2009a)
due to the fact that they assumed constant brightness temperature when
extrapolating to CO 1-0 from the observed 4-3 luminosity. In the
following section we show that the 4-3 line is subthermally excited.

The average gas surface density over the disk is 2400 \msun pc$^{-2}$,
with some (resolved) knots a factor two or so higher. Tacconi et
al. (2008) find average values in $z \sim 2$ SMGs of $\sim 5000 \msun$
pc$^{-2}$, while maximum gas surface densities of 10,000 \msun
pc$^{-2}$ are seen in the ULIRGs at low redshift (Wilson et al.  2009).
For comparison, the canonical values for gas surface densities for
Galactic GMCs are $\sim 170$ \msun pc$^{-2}$ on scales $\sim 50$pc, and
only get as high as 10,000 \msun pc$^{-2}$ on scales approaching those
of dense molecular cores ($\le 1$ pc), directly associated
with star formation (Solomon et al. 1987).  Gas surface densities
as high as  10,000 \msun pc$^{-2}$ have also been seen in the
host galaxies of high redshift quasars (Riechers et al. 2009).

The ratio of total IR to L$'_{\rm CO 1-0}$ luminosity in GN20 is 181
L$_\odot$ (K \kms pc$^{2}$)$^{-1}$.  For comparison, nearby spirals
typically have ratios between 20 and 100, while nearby low redshift
ULIRGs, $z \sim 2$ SMGs, and high redshift quasar host galaxies have
ratios between 100 and 1000 (eg. Gao \& Solomon 2004; Solomon \&
Vanden Bout 2005; Carilli et al. 2002; Wang et al. 2009; Daddi et al.
2010b). GN20 falls within the scatter of the FIR -- L'CO correlation
as presented in eg. Solomon \& Vanden Bout (2005).

The dust mass is $8 \times 10^8$ \msun, derived from the SED fitting,
and assuming $\kappa_{125\mu m} = 18.75$ cm$^2$ g$^{-1}$ (Hildebrand
1983).  The gas to dust mass ratio in GN20 is $190 \times
(\alpha/0.8)$. This is somewhat higher than the mean value for low redshift
ULIRGs, of $120 \pm 28$ derived by Wilson et
al. (2009), who also quote a typical value in the Milky Way of 150.
However, we emphasize that there are significant uncertainties in both
$\alpha$ and the dust absorption coefficient when deriving the value
for GN20.

\subsection{CO excitation}

Figure 8 shows the CO excitation ladder for GN20.  We have fit these
data with a standard radiative transfer, large velocity gradient (LVG)
model (Scoville \& Solomon 1974), ie. a model where photon trapping
can be considered strictly a local phenomenon.  We use the collision
rates from Flower et al. (2001) with an ortho-to-para ratio of 3 and a
CO abundance per velocity gradient of [CO]/$\Delta v = 1\times
10^{-5}\,{\rm pc}$ (km s$^{-1}$)$^{-1}$.  The observed line flux
densities were compared to the LVG predicted line brightness
temperatures as in Weiss \etal\ (2007). The analysis thus yields not
only the gas density and temperature, but also an estimate of the CO
source solid angle, $\Omega_{\rm S}$, which can be expressed in terms
of the equivalent radius $r_{0}= D_{A}\,\sqrt{\Omega_{\rm S}/\pi}$,
i.e.  the source radius if the CO were distributed in a face-on,
filled circular disk.

We find that the data are poorly described by a single component fit,
under-predicting by a factor two the observed CO 1-0 luminosity.  The
data are reasonably fit by a two component gas model, including a
diffuse, lower excitation component and a more concentrated, higher
excitation gas (Figure 8a).  For the lower excitation component we
constrain the source radius to be 4.5 kpc, based on the CO 2-1 imaging
(Section 4.3).  For this component, we derive a filling factor of 0.5,
an H$_2$ density of 300 cm$^{-3}$, and a kinetic temperature of
30K. The low density is required to ensure that the diffuse gas does
not contribute significantly to the J=4-3 or higher CO transitions.
For the higher excitation component, we fix the source radius to be
2.5 kpc, based on the CO 6-5 observations (Section 4.3).  We then
derive a density of 6300 cm$^{-3}$, a kinetic temperature of 45 K, and
a filling factor of 0.13.  The low excitation component contributes
about half of the 1-0 emission, with a decreasing contribution to
higher order emission. If we assume the same CO 1-0 to H$_2$ mass
conversion factor, this would imply about half the gas mass is in the
more extended, low excitation component, and the other half is in the
more compact component.  We can speculate that the lower excitation
component may have a higher conversion factor, more typical of the
Milky Way, in which case its gas mass would be a factor five
higher. However, such a high mass may violate the dynamical
constraints below.

We have compared our two component CO excitation model for GN20 to the
inner disk of the Milky Way, the starburst regions of NGC 253 (inner
few hundred parsecs), and the FIR hyper-luminous ($L_{FIR} \sim
10^{13}$ L$_\odot$) host galaxy of the $z = 4.7$ quasar BR1202-0725
(see Weiss et al. 2005; Riechers et al. 2006 for details). We note
that the Milky Way excitation also holds for the CO excitation in
normal star forming galaxies at $z \sim 2$ (Dannerbauer et al. 2009;
Aravena et al. 2010).  Figure 8b shows the fit CO ladder for all the
sources normalized to CO 1-0. The excitation of the integrated CO
emission from GN20 is higher than the Milky Way, but much lower than
high redshift quasar hosts and the nuclear starburst regions of nearby
galaxies.

The integrated CO ladder from GN20 implies that the CO 4-3 line
strength is more than a factor two lower than expected for thermal
excitation, due to a major contribution from the low excitation
component at low order. Hence one should use caution before deriving
total H$_2$ masses by extrapolating to lower order transitions based
on observations of CO 3-2 or higher, and assuming thermal
excitation. For example, Tacconi et al. (2006) find typical densities
in SMGs at $z \sim 2$ to be $> 1000$ cm$^{-3}$. However, they base
this conclusion on LVG fits to CO 3-2 and higher order transitions.
The diffuse component in GN20 only becomes dominant in the 2-1 and 1-0
transitions, ie. the Tacconi et al. study is relevant to the compact
component for the CO, but can say little about the full molecular gas
reservoirs.

\subsection{Dynamical mass and timescales}

The CO 6-5 velocity channel maps, and moment maps, are shown in Figure
4 and 5. These show a regular velocity gradient across the disk. We
interpret this as a rotating disk with an inclination angle $\sim
45^o$, based on Figure 5b, although we emphasize that this estimate is
uncertain. From the channel images, the total observed velocity
gradient corresponds to 750 km s$^{-1}$ over $\sim 1.0"\pm
0.2"$ (section 4.2 and 4.3). This implies a rotational velocity $\sim
570$ km s$^{-1}$ over $\sim 4$ kpc radius. The dynamical mass is
then $3 \times 10^{11}$ M$_\odot$.

Admittedly, these numbers are approximate, and the area for the total
CO mass may be a bit larger than that over which the rotation curve is
measured (Section 4.4). However, it is clear that the gas mass
contributes substantially ($\sim 40\% \times (\alpha/0.8$)) to the
total mass of the system within 4 kpc radius. Indeed, there is little
room for a much larger CO-to-H$_2$ conversion factor.  This high
percentage is comparable to the typical value found for $z \sim 2$
SMGs (Tacconi et al. 2006), as well as comparable to what has been
found for more normal forming galaxies at $z \sim 2$ (Tacconi et
al. 2010; Daddi et al. 2010a), but is considerably larger than the
$\sim 16\%$ gas fractions seen in nearby ULIRGs. The combined stellar
plus gas mass is comparable to (in fact, slightly larger than), the
dynamical mass, although, again, the stellar (and gas) masses are
derived over a larger area than sampled by the dynamics of the CO 6-5
emission, and there remains the uncertainties of the inclination
angle, gas conversion factor, and optical obscuration.

Using the gas mass and star formation rate, the gas consumption
timescale ($\equiv$ gas mass/SFR) for GN20 is $\sim 5\times 10^7
\times (\alpha/0.8$) years. The rotational time for the disk is also
$\sim 5\times 10^7$ years. Hence, the gas consumption timescale is
comparable to the dynamical timescale in GN20.  The $z \sim 2$ SMG
sample of Tacconi et al. (2006) have comparable gas consumption
timescales to GN20.  This compares to the order-of-magnitude longer
gas consumption timescales found for normal star forming galaxies at
$z \sim 2$ (Tacconi et al. 2010; Daddi et al. 2010a).

\section{Discussion}

\subsection{GN20} 

We have presented the most detailed imaging analysis to date of the CO
emission from an SMG, including imaging the lower order
transitions down to 1 kpc resolution. The principle physical
parameters resulting from this study are listed in Table 1.

The main result from this work is that the molecular gas and star
formation are well resolved on a scale $\sim 10$kpc. The high
resolution CO 2-1 imaging, in particular, shows a partial ring, or
disk, on this scale.  The ring shows a few resolved clumps with
(deconvolved) sizes $\sim 2$ kpc, but no single clump dominates the
total emission. This is also true for the 1.4 GHz continuum emission,
presumably tracing star formation, which has a similar morphology to
the CO emission.  The higher order CO observations indicate a regular
velocity field, consistent with a disk with a rotational velocity of
570 km s$^{-1}$.  The total H$_2$ mass derived from the CO 1-0
emission is $1.3\times 10^{11} \times (\alpha/0.8)\msun$, which is
roughly 40\% of the dynamical mass within 4 kpc radius.

The entire $\sim 10$ kpc region of active star formation, as traced by
the CO, FIR, and radio continuum, is completely obscured in the HST
I-band (rest-frame UV) image.

The CO is lower excitation than seen in low redshift nuclear
starbursts and high redshift quasar host galaxies, but it is higher
than in nearby spiral galaxies and normal star forming galaxies at $z
\sim 1.5$.  The CO emission from GN20 is consistent with a two
component model, consisting of a 4.5 kpc radius disk of lower density
(300 cm$^{-3}$), temperature (30K), with a filling factor $\sim 0.5$,
and a region of $\sim 2.5$ kpc radius with higher density ($\sim 6300$
cm$^{-3}$), higher temperature (45K), and lower filling factor ($\sim
0.13$). The mass is roughly equal in each component (assuming the same
conversion factor). The gas depletion timescale is comparable to the
rotational time of the galaxy $\sim 5 \times 10^7 \times (\alpha/0.8$)
years. We note that Papadopoulos et al. (2010) have proposed that dust
opacity in dense regions can also affect the observed line ratios for
CO, when observing very high order transitions (eg. CO6-5). High
resolution imaging of the CO6-5 is required to determine the spatial
dependence of gas excitation in GN20.
 
\subsection{Star formation in GN20}

Two mechanisms have been proposed in recent years for driving active
star formation in high redshift galaxies: major gas rich mergers
(Narayanan et al. 2009) and cold mode accretion (Dekel et al. 2009;
Keres et al. 2009; Keres et al. 2005). The process of fueling nuclear
starbursts via major gas rich mergers is well studied in the nearby
Universe (eg. Mihos \& Hernquist 1996; Barnes \& Hernquist 1991). The
general idea is that gravitational torques induce strong dissipation
and inflow of gas, leading to an increase in the star formation rates
by up to two orders of magnitude over quiescent disks on the short
timescale of the merger $\sim$ few$\times 10^7$ years.  Most of this
star formation occurs on scales $< 1$ kpc in the galaxy nuclei, as is
seen in nearby ULIRGs (Downes \& Solomon 1998). Major gas rich mergers
have been invoked to explain the compact, maximal starbursts seen in
some high redshift quasar host galaxies (Li et al. 2007; Walter et
al. 2009; Riechers et al. 2009). Johansson et al. (2009) point out, if
a nuclear starburst is merger-driven, the mass ratio has to be close
to unity.

Tacconi et al. (2006; 2008) conclude, based on high resolution CO
imaging, that $z \sim 2$ SMGs are nuclear starbursts on scales
$<4$kpc, driven by major gas rich mergers. However, they base this
conclusion on observations of high order CO lines (3-2 and higher at
sub-arcsecond resolution). In GN20 we see a more extended, lower
excitation molecular gas distribution on a scale $\sim 10$ kpc,
containing at least half the gas mass in the system.  Interestingly, a
number of other $z \sim 2$ SMGs have been observed in CO 1-0: the
submm-bright ERO J16450+4626 (Greve et al. 2003), SMM J13120+4242
(Hainline et al.  2006), and SMM J02399-0136 (Ivison et al. 2010).
These galaxies show excess CO 1-0 emission relative to what is
expected by extrapolating from higher-order transitions assuming
constant brightness temperature, by factors of at least two.
Moreover, VLA imaging of J16450+4626 reveals extended CO 1-0 emission
on a scale of $\sim 10$ kpc (Greve et al. 2003), while for J02399-0136
the CO 1-0 emission extends over 25 kpc (Ivison et al. 2010).

An alternative model, known as cold mode accretion (CMA), or stream
fed galaxy formation, has recently been proposed to explain secular
star formation (ie. on timescales $> 10^8$ years) in more populous,
normal star forming galaxies at $z \sim 2$ (Dekel et al. 2009; Keres
et al. 2009).  In the CMA model, gas flows into galaxies from the IGM
along cool, dense filaments. The flow never shock-heats due to the
rapid cooling time, but continuously streams onto the galaxy at close
to the free-fall time. This gas forms a thick, turbulent, rotating
disk which efficiently forms stars across the disk, punctuated by
giant clouds of enhanced star formation on scales $\sim$ few kpc.
These star forming regions then migrate to the galaxy center via
dynamical friction and viscosity, forming compact stellar bulges
(Genzel et al. 2006; Genzel et al. 2008; Bournaud et al.  2008a,b;
Elmegreen et al. 2009).  The CMA process can lead to relatively steady
and active ($\sim 100$ M$_\odot$ yr$^{-1}$) star formation in galaxies
over timescales approaching 1 Gyr.  The process slows down
dramatically as gas supply decreases, and the halo mass increases,
generating a virial shock in the accreting gas.  Subsequent dry
mergers at lower redshift then lead to continued total mass build up,
and morphological evolution, but little subsequent star formation
(Hopkins et al. 2009; Naab et al. 2009).  Observations of intermediate
redshift ($z \sim 2$), normal star forming galaxies support the CMA
model (Genzel et al. 2006, 2008; Daddi et al. 2008; 2009b; 2010a;
Tacconi et al. 2010).

Dav\'e et al (2009) suggest that a substantial fraction of SMGs could
be fed primarily by CMA, and not major mergers.  Hydrodyanamic
simulations show that quite high gas accretion rates can be achieved
in large halos at early epochs, and elevations of SFR over the average
accretion rate by a factor 2 to 3 can be frequent owing to (common)
minor mergers.  While the Dav\'e et al. study focused on $z=2$,
Finlator et al. (2006) studied galaxies in cosmological hydrodynamic
simulations at $z=4$ and found two instances (within a $3\times 10^6$
comoving Mpc$^3$ volume) of galaxies forming stars at $>1200$
M$_\odot$ yr$^{-1}$.  These galaxies had stellar masses
several~$\times 10^{11}M_\odot$, very similar to GN20.  They were
found to be forming stars at $\sim 2$ to $2.5\times$ their average
rate, owing to their location at the center of the largest potential
wells with constantly infalling satellite galaxies, ie.  environments
comparable to GN20.  While their star formation rates are still a
factor $\sim 2$ to 3 lower than GN20, given the uncertainties in
conversion from $L_{IR}$ to SFR (eg.  the IMF), and uncertainties in
the models, there is at least a plausible association of these
simulated galaxies with GN20.

The CMA and major merger scenarios might be expected to have different
structural and kinematic signatures.  The ordered rotation and
extended gas distribution would favor a disk that is not being
strongly disturbed.  The gas distribution (Figure 2) and velocity
field (Figure 5) of GN20 are qualitatively similar to that seen in the
simulated SMG maps in Figure~5 of Dav\'e et al. (2009), particularly
object ``B" which is a quiescently star-forming thick disk, though we
note that the SFR of this $z=2$ simulated galaxy is substantially
lower than GN20. On the other hand, such signatures do not
conclusively rule out a merger, as Robertson \& Bullock (2008) have
shown that ordered rotation of an extended gas disk can be
reestablished very shortly after a major merger.  Therefore we cannot
make firm conclusions about the driver of star formation in GN20,
although the simplest interpretation of the molecular gas data favors
CMA.  A key point is that, if GN20 is an on-going major gas rich
merger, then some process has managed to ensure that the star
formation and molecular gas distribution has not been focused into one
or two compact nuclear regions. 

This latter point also begs the question of the progeny of a system
such as GN20, since massive (stellar masses $\sim 10^{11}$ M$_\odot$)
'red and dead' ellipticals at $z \sim 2$ typically show fairly compact
stellar distributions, with radii $\sim 1$kpc (van Dokkum et
al. 2008), although questions have been raised about morphological
K-corrections and the presense of an AGN (Daddi et al. 2005). On the
other hand, star forming galaxies (SFR $\sim 100$ M$_\odot$
year$^{-1}$) of this stellar mass at $z \sim 2$ are often seen to have
spatially extended stellar distributions, comparable to the CO size of
GN20 (Kriek et al. 2009; Daddi et al. 2008; 2010a). Hence, if GN20 is
to evolve into a passive elliptical at $z \sim 2$, the stellar
distribution will have to evolve to a more compact configuration.
Conversely, GN20 could remain a star forming galaxy for a long period,
although at a substantially lower star formation rate.

A number of key observations are required to untangle the mechanisms
driving star formation in GN20. First, high resolution imaging of the
millimeter continuum will reveal the distribution of star formation
across the disk in greater detail.  Second, high resolution imaging of
the high order transitions can be used to determine the spatial
excitation of the molecular gas. And third, high spectral and spatial
resolution imaging of the low order CO emission is required to
determine the gas dynamics on both large and small scales, ie. verify
the overall rotation, and determine the internal turbulent velocity
and stability parameters in the disk. Shapiro et al. (2009) show that
such a study of disk kinemetry enables an 'empirical differentiation
between merging and non-merging systems'.  These latter observations
have now become possible with the Expanded Very Large Array.

\subsection{Lensing}

An open issue remains gravitational lensing, in particular given the
partial ring-like morphology of GN20 in CO emission.  An Einstein ring
has been observed in CO in the strongly lensed, $z = 4.12$ quasar host
galaxy J2322+1944 (Carilli et al. 2003; Riechers et al. 2008). Lensing
would also help explain the extreme (apparent) luminosity of GN20
(Pope et al. 2006).

To date, there is no evidence for a lensing galaxy in the HST image of
GN20. This is unusual, given the depth of the HST data, and the $\sim
1''$ diameter of the ring. For example, the CO Einstein ring in
J2322+1944 has a similar diameter as GN20, and the lensing galaxy is
seen clearly in the F814W filter image with a magnitude of 21.9.  This
galaxy would be easily detected in the HST I-band image of GN20.
Also, the regular velocity field in the 6-5 line argues against
lensing, since caustic structures can lead to complex apparent
velocity structures in the image plane, as is the case for J2322+1944
(Riechers et al. 2008). Lastly, the fact that the gas mass and the
dynamical mass are comparable (within a factor 2 or so), argues
against very strong lensing.

\acknowledgments
 
CC thanks the Max-Planck-Gesellschaft and the Humboldt-Stiftung for
support through the Max-Planck-Forschungspreis, and the Max-Planck
Institute for Astronomie in Heidelberg for their hospitality. ED
acknowledges the funding support of an ERC starting resaerch grant and
French ANR under contracts ANR-07-0228 and ANR-08-JCJC-0008.  DR
acknowledges support from from NASA through Hubble Fellowship grant
HST-HF-51235.01 awarded by the Space Telescope Science Institute,
which is operated by the Association of Universities for Research in
Astronomy, Inc., for NASA, under contract NAS 5-26555. We thank
T. Muxlow and J. Younger for sharing their data, and Christian Henkel
for his work on the LVG code. We thank the referee for useful
comments.


\clearpage
\newpage

\begin{table}\label{Observed}
\begin{center}
\caption{Observed and derived results for GN20}
\begin{tabular}[ht]{|c|c|}
\tableline
Source & J123711.89+622211.7$^a$ \\
S$_{1.4GHz}$ & $72\pm 13\mu$Jy \\
S$_{23GHz}^b$ & $< 60\mu$Jy \\
S$_{43GHz}^c$ &  $< 174\mu$Jy \\
S$_{137 GHz}$ & $0.89 \pm 0.15$mJy \\
L$_{FIR}^d$ & $2.9\times 10^{13}$ \lsun \\
SFR$^e$ & 3000 \msun yr$^{-1}$ \\
S(CO 6-5) &  $2.7 \pm 0.3$ mJy \\
FWHM (CO 6-5)& $664 \pm 50$ km s$^{-1}$ \\
I(CO 1-0) & $0.21\pm0.05$ Jy km s$^{-1}$ \\
I(CO 2-1) & $0.64 \pm 0.16$ Jy km s$^{-1}$ \\
I(CO 4-3)$^f$ & $1.5 \pm 0.2$ Jy km s$^{-1}$ \\
I(CO 5-4) & $2.2 \pm 0.7$ Jy km s$^{-1}$ \\
I(CO 6-5) & $1.8\pm 0.2$ Jy km s$^{-1}$ \\
I(CO 7-6)$^g$ & $< 1.2$ Jy km s$^{-1}$ \\
L(CO 1-0) & $7.8\times 10^6$ \lsun  \\
L$'_{\rm CO 1-0}$ & $1.6\pm 0.04 \times 
10^{11}$ K km s$^{-1}$ pc$^{-2}$ \\
M(H$_2$) & $1.3\times 10^{11} \times (\alpha/0.8)$ \msun \\
M$_{dyn}^h$ & $3.0\times 10^{11}$ \msun \\
M$_{stellar}^i$ & $2.3\times 10^{11}$ \msun \\
\tableline
\end{tabular}
\end{center}
\noindent $^a$Peak position at 1.4GHz at 1.7$"$ resolution. \\
\noindent $^b$Continuum 2$\sigma$ limit at 23 GHz at 3.5$"$ resolution. \\
\noindent $^c$Continuum 2$\sigma$ limit at 43 GHz at 0.5$"$ resolution. \\
\noindent $^d$The total IR luminosity inferred from SED fitting (Daddi et al.
2009a). \\
\noindent $^e$Total star formation rate derived from the SED assuming
a Chabrier (2003) IMF. \\
$^f$From Daddi et al. (2009a). \\
$^g$2$\sigma$ limit from Casey et al. (2009). \\
$^h$Dynamical mass inside a radius of 4 kpc (Section 5.5) \\
$^i$Stellar mass derived by Daddi et al. (2009a). \\ 

\end{table}

\clearpage
\newpage

\noindent  Alexander, D., Bauer, F., Chapman, S. et al. 2005, ApJ,
  632, 736

\noindent  Aravena, M., Bertoldi, F., Carilli, C. et al. 2009, ApJL,
in press (arXiv:0911.4297)

\noindent  Aravena, M., Carilli, C., Daddi, E. et al. 2010, ApJ, submitted

\noindent  Barnes, J. \& Hernquist, L. 1991, ApJ, 370, 65

\noindent  Bigiel, F., Walter, F., Brinks, E. et al. 2008, AJ, 136, 2846

\noindent  Blain, A., Smail, I, Ivison, R., Kneib, J.-P., Frayer, D. 
2002, PhR 369 111


\noindent  Bournaud, F., Daddi, E., Elmegreen, B. et al. 2008a, A\&
  A, 486, 741

\noindent  Bournaud, F., Elmegreen, B., Elmegreen, D. 2008b, ApJ,
670, 237

\noindent  Cornwell, T., Braun, R., Briggs, D. 1999, in 'Synthesis
Imaging in Radio Astronomy II,' eds. G. B. Taylor, C. L. Carilli, \&
R. A. Perley. ASP Conference Series, Vol. 180, 151

\noindent  Carilli, C.L. \& Holdaway, M. 1999, Radio Science, 34, 817

\noindent  Carilli C.L. \& Yun, M.S. 2000, ApJ, 530, 618

\noindent  Carilli, C.L., Lewis, G.F., Djorgovski, S.G. et al. 2003,
Science, 300, 77

\noindent  Carilli, C.L, Kohno, K, Kawabe, R. et al. 2002, AJ, 123,
1838

\noindent  Casey, C.M., Chapman, S., Daddi, E. et al. 2009, MNRAS, in press

\noindent  Chabrier, G. 2003, PASP, 115, 763

\noindent  Chapman, S., Blain, A., Ivison, R., Smail, I. 2003 Nature
  422 695

\noindent  Chapman, S., Blain, A., Ibata, R. et al. 2009, ApJ, 691, 560

\noindent  Cimatti, A., Daddi, E., Renzini, A. et al. 2004, Nature,
  430, 184

\noindent  Collins, D., Stott, J.P., Hilton, M. et al. 2009, Nature,
458, 603

\noindent  Coppin, K.E., Smail, I., Alexander, D. et al. 2009,
MNRAS, 395, 1905


\noindent Daddi, E., Renzini, A., Pirzkal, N. et al. 2005, ApJ, 626,
680

\noindent  Daddi, E., Dickinson, M., Morrison, G. et al 2007, ApJ
  670 156

\noindent  Daddi, E., Dannerbauer, H., Elbaz, D. et al. 
2008, ApJ 673 L21

\noindent  Daddi, E., Dannerbauer, H., Stern, D. et al. 
2009a, ApJ, 694, 1517

\noindent  Daddi, E., Dannerbauer, H., Krips, M. et al. 
2009b, ApJ, 695, L176

\noindent  Daddi, E., Bournaud, F., Walter, F. et al. 
2010a, ApJ in press (arXiv:0911.2776)

\noindent  Daddi, E., Elbaz, D., Walter, F. et al. 2010b, 
ApJL, submitted

\noindent  Dannerbauer,H., Lehnert, M., Lutz, D.
et al. 2002 ApJ 573 473

\noindent  Dannerbauer, H., Lehnert, M., Lutz, D. et al.
2004, ApJ, 606, 664

\noindent  Dannerbauer, H., Walter, F., Morrison, G. 
2008, ApJ, 673, L127

\noindent  Dannerbauer, H., Daddi, E., Riechers, D. 
et al. 2009, ApJ, 698, L178

\noindent  Dav\'e, Romeel, Finlator, K., Oppenheimer, B. et al.
2009, MNRAS, in press  (arXiv0909.4078)

\noindent  Dekel, A., Birnboim, Y., Engel, G. et al. Nature 2009,
  457, 451

\noindent  Downes \& Solomon 1998 ApJ, 507, 615

\noindent  Doherty, M., Tanaka, M., de Breuck, C. et al.
2009, A\& A, in press

\noindent  Dunlop, J., McCure, R., Yamada, T. 
et al. 2004 MNRAS 350 769

\noindent  Elmegreen, B.G., Elmegreen, D.M., Fernandex, M.X.,
Lemonias, J.J. 2009, ApJ, 692, 12

\noindent  Finlator, K., Dav\'e, R., Papovich, C., \& Hernquist,
  L. 2006, ApJ, 639, 672

\noindent  Flower, D.R. 2001, J. Phys. B; At. Mol. Opt. Phys. 34, 1

\noindent  Gao, Y., \& Solomon, P. 2004, ApJS, 152, 63

\noindent  Greve, T., Ivison, R., Papadopolous, P. 2003, ApJ, 599,
  839

\noindent  Genzel, R. Tacconi, L. J., Eisenhauer, F. et al. 2006,
  Nature, 442, 786

\noindent  Genzel, R., Burkert, A., Bouché, N. et al. 2008, ApJ,
  687, 59

\noindent  Giavalisco, M., Dickinson, M., Ferguson, H. et al. 2004,
ApJ, 600, L103

\noindent  Hainline, L., Blain, A., Greve, T. et al. 2006, ApJ, 650,
614

\noindent  Hildebrand, R.H. 1983, QJRAS, 24, 267

\noindent Hopkins, P., Bundy, K., Murray, N. et al. 2009, MNRAS,
398, 898

\noindent  Iono, D., Peck, A., Pope, A. et al. 2006, ApJ, 640, L1

\noindent  Ivison, R., Smail, I., Papadopoulos, P. et al.
2010, MNRAS, in press (arXiv:0912.1591)

\noindent Johansson, P, Naab, T., Burkert, A. 2009, ApJ, 690, 802

\noindent  Keres, D., Katz, N., Fardal, M.  et al. 2009, MNRAS, 395, 160
 qq
\noindent  Keres, D., Katz, N., Weinberg. D., Dav\'e, R. 2005,
MNRAS, 363, 2

\noindent  Knudsen, K., Kneib, J.-P., Richard, J.,
Petitpas, G., \& Egami, E. 2009, ApJ, in press

\noindent  Kotilainen, J., Falomo, R., Decarli, R. et al.
2009, ApJ, 703, 1663

\noindent  Kurk, J., Cimatti, A., Zamorani, G. et al. 2009,
A\& A, 504, 331

\noindent  Kriek, M., van der Wel, A., van Dokkum, P. et al. 2008,
ApJ, 682, 896 

\noindent  Kriek, M., van Dokkum, P., Franx, M. et al. 2009,
ApJ, 705, L71

\noindent  Leroy, A., Walter, F., Brinks, E. et al. 2008,
AJ, 136, 2782

\noindent  Li, Y., Hernquist, L., Robertson, B. et al. 2007, ApJ,
  665, 187

\noindent  Mihos, C. \& Hernquist, L. 1996, ApJ, 464, 641

\noindent  Miley, G., \& de Breuck, C. 2007, A\& A Rev. 15, 67

\noindent  Mobasher, B., Dickinson, M., Ferguson, H. et al.
2005, ApJ, 635, 832

\noindent  Momjian, E., Riechers, D., 
Carilli, C. et al. 2009, AJ, in press (arXiv:1002.3324)

\noindent  Momjian, E., Carilli, C. \& Petric, A. 2005, AJ, 129, 1809

\noindent  Morrison, G. et al. 2009, ApJ, submitted

\noindent Naab, T., Johansson, P., Ostriker, J. 2009, ApJ, 699, L178

\noindent  Noeske, K.G., Weiner, B.J., Faber, S.M. et al. 
2007, ApJ, 660, L43

\noindent  Noeske, K.G., Faber, S.M. Weiner, B.J. et al. 
2007, ApJ, 660, L47

\noindent  Papadopoulos, P., Isaak, K., van der Werf,P. 2010,
ApJ, submitted (arXiv:1001.3653)

\noindent  Pannella, M., Carilli, C.L., Daddi, E. et al.
2009, ApJ, 698, L16

\noindent  Papovich, C., Momcheva, I., Willmer, C. et al.
2010, ApJ, submitted (arXiv:1002.3158)

\noindent  Pope, A., Scott, D., Dickinson, M. 
et al. 2006 MNRAS 370 1185

\noindent  Renzini, A. 2006, ARAA, 44, 141


\noindent  Riechers, D., Walter, F., Carilli, C. et al. 
2006, ApJ, 650, 604

\noindent  Riechers, D., Walter, F., Brewer, B. 
2008, ApJ, 686, 851

\noindent  Robertson, B. \& Bullock, J. 2008, ApJ, 685, L27

\noindent  Riechers, D., Walter, F., Carilli, C. 2009, ApJ, 703,
  1338

\noindent  Shapiro, K., Genzel, R., F\"orster-Schreiber, N. et
al. 2009, ApJ, 682, 231

\noindent  Schinnerer, E., Carilli, C.L., Capak, P. et al. 2008,
ApJ, 689, L5

\noindent  Schinnerer, E., Capak, P., Carilli, C.L. et al. 2009,
submitted to A\& A

\noindent  Scoville, N.  \& Solomon, P. 1974, ApJ, 187, 67L

\noindent  Solomon \& Vanden Bout 2005, ARAA, 43, 677

\noindent  Solomon, P., Rivolo, A., Barrett, J., \& Yahil, A. 1987,
  319, 730

\noindent  Stevens, J, Ivison, R., Dunlop, J. 
et al. 2003 Nature, 425, 264

\noindent  Tacconi, L., Neri, R., Chapman, S. et al. 2006, ApJ, 640, 228

\noindent  Tacconi, L., Genzel, R., Smail, I., et al. 2008, ApJ, 680, 246

\noindent  Tacconi, L., Genzel, R., Neri, R. et al. 2010, Nature, in press

\noindent  Thompson, T. 2008, ApJ, 684, 212

\noindent  Walter, F, Riechers, D., Cox, P. et al. 2009,
Nature, 457, 699

\noindent  Wang et al. 2007 ApJ 670 L89

\noindent  Wang, R., Carilli, C., Wagg, J, et al. 2009, ApJ,
  submitted

\noindent  Wagg, J., Owen, F., Bertoldi, F. et al. 2009, 
ApJ, 699, 1843

\noindent  Weiss, A., Walter, F., \& Scoville, N. 2005, A\& A, 
438, 533

\noindent  Weiss, A., Downes, D., Walter, F., Henkel, C. 2007,
ASPC Vol. 375, p. 25

\noindent  Wiklind, T., Dickinson, M., Ferguson, H. et al.  2008,
  ApJ, 676, 781

\noindent  Wilson, C.D. et al. 2009, ApJS, 178, 189

\noindent  Younger, J., Fazio, G., Huang, J.S. et al. 2007, ApJ,
  671, 1531

\noindent  Younger, J., Fazio, G., Wilner, D. et al. 2008, 
ApJ, 688, 59

\noindent  Yun, M.S., Reddy, N., Condon, J. 2001, ApJ, 554, 803

\noindent  Zheng, X.Z., Bell, E.F., Papovich, C. et al. 2007,
ApJ, 661, L1

\clearpage
\newpage
 
\begin{figure}
\psfig{file=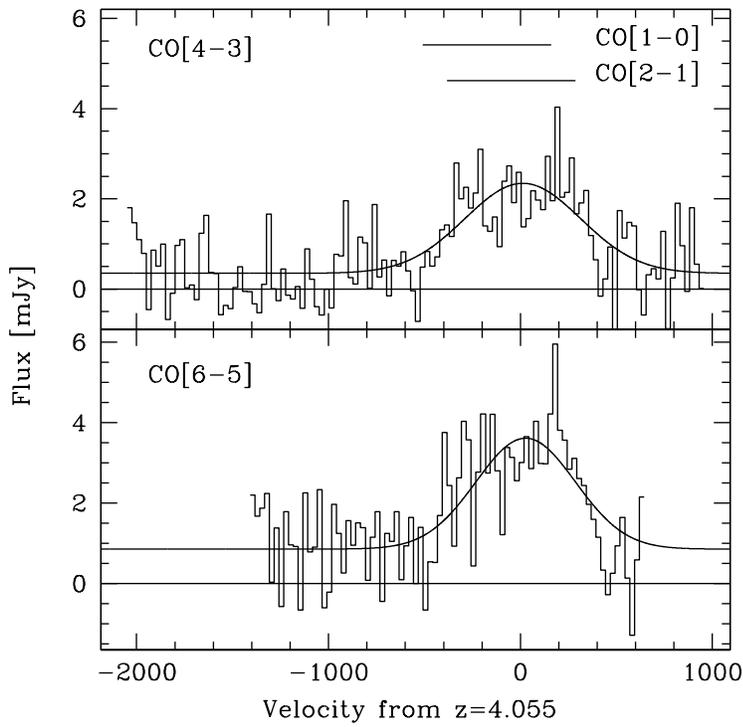,width=4in}
\caption{The CO 6-5 and CO 4-3 spectra of GN20 from the Plateau de
Bure Interferometer, along with a Gaussian fit
with parameters given in Section 4.3.  The spectral
coverage for the observed VLA bands in CO 2-1 and CO 1-0 are shown. 
}
\end{figure}

\begin{figure}
\psfig{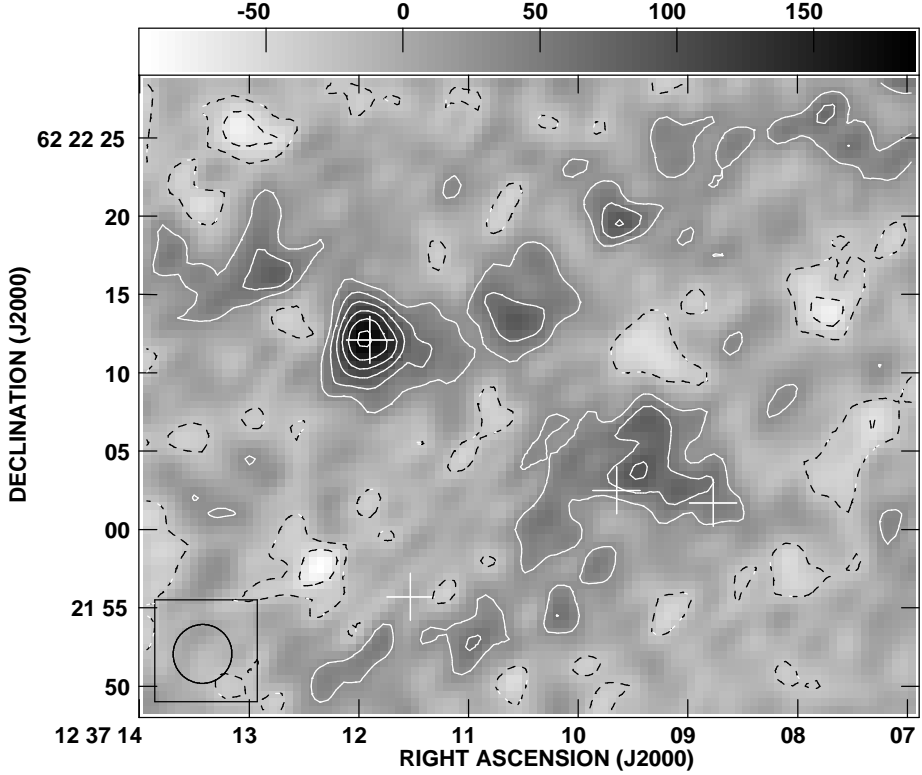}
\caption{
The VLA image of the CO 1-0 emission line from GN20 using
the D-array data, with a circular Gaussian restoring beam of
FWHM = $3.7"$. Data is from a single 50 MHz IF centered at  
22.815 GHz.  The crosses show the optical positions of 
GN20, GN20.2b, GN20.2a, and a $z =4.05$ LBG, from north to south.
Contour levels are: -60, -30, 30, 60, 90, 120, 150, 
180 $\mu$Jy beam$^{-1}$, the rms on the image is
30 $\mu$Jy beam$^{-1}$, and the greyscale flux units are in $\mu$Jy 
beam$^{-1}$. 
}
\end{figure}

\begin{figure}
\psfig{file=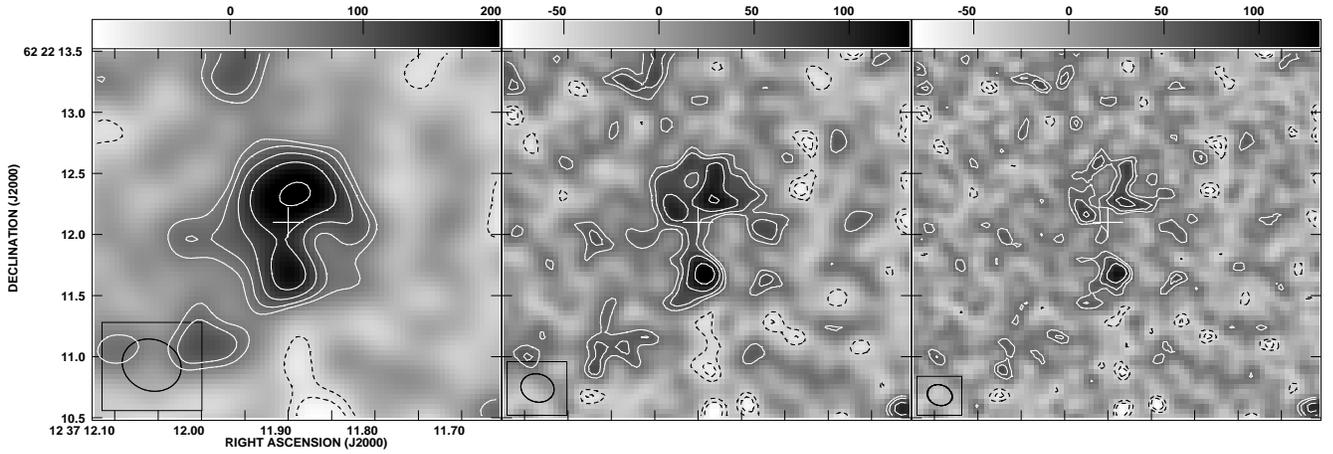,width=7in}
\caption{
The VLA images of the CO 2-1 emission line from GN20, using the B, C,
D-array data, using different Gaussian tapers of the visibilities as
described in Section 4.2.  The data is from two 50~MHz IFs centered at
45.585 and 45.635 GHz.  The cross in this, and subsequent images,
shows the radio peak position of GN20 at 1.7$"$ resolution. The
contour levels are a geometric progress in square root two, starting
at 55$\mu$Jy beam$^{-1}$ (left at $0.45"$ resolution), and 42$\mu$Jy
beam$^{-1}$ (middle at $0.25"$ resolution, right at $0.18"$
resolution), such that two contour levels implies a factor two change 
in surface brightness.  Negative contours are dashed. The rms in the
three images are: 37, 27, and 26 $\mu$Jy beam$^{-1}$, respectively.
}
\end{figure}

\begin{figure}
\psfig{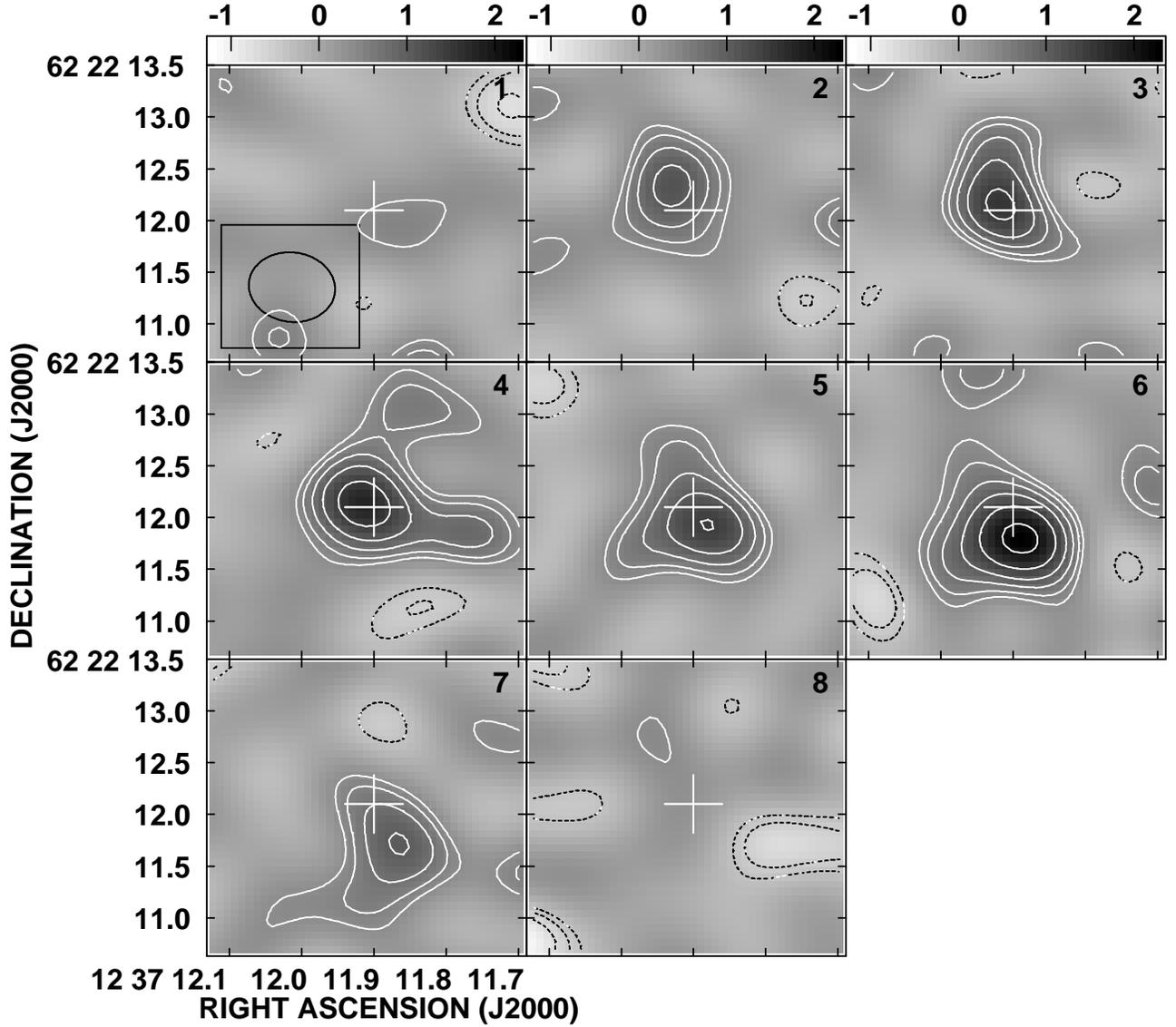}
\caption{
The CO 6-5 velocity channel maps at 150 km s$^{-1}$ channel$^{-1}$\ ,
and $0.84'' \times 0.67''$ resolution (UN weighting).  
Contour levels are a 
geometric progression in the square root two starting at 
0.36 mJy beam$^{-1}$. Negative contours are dashed. 
The greyscale flux units are in mJy 
beam$^{-1}$. The rms in each channel is 0.25 mJy beam$^{-1}$.
}
\end{figure}

\begin{figure}
\psfig{file=fig5a.ps,width=4in}
\vskip -0.3in
\psfig{file=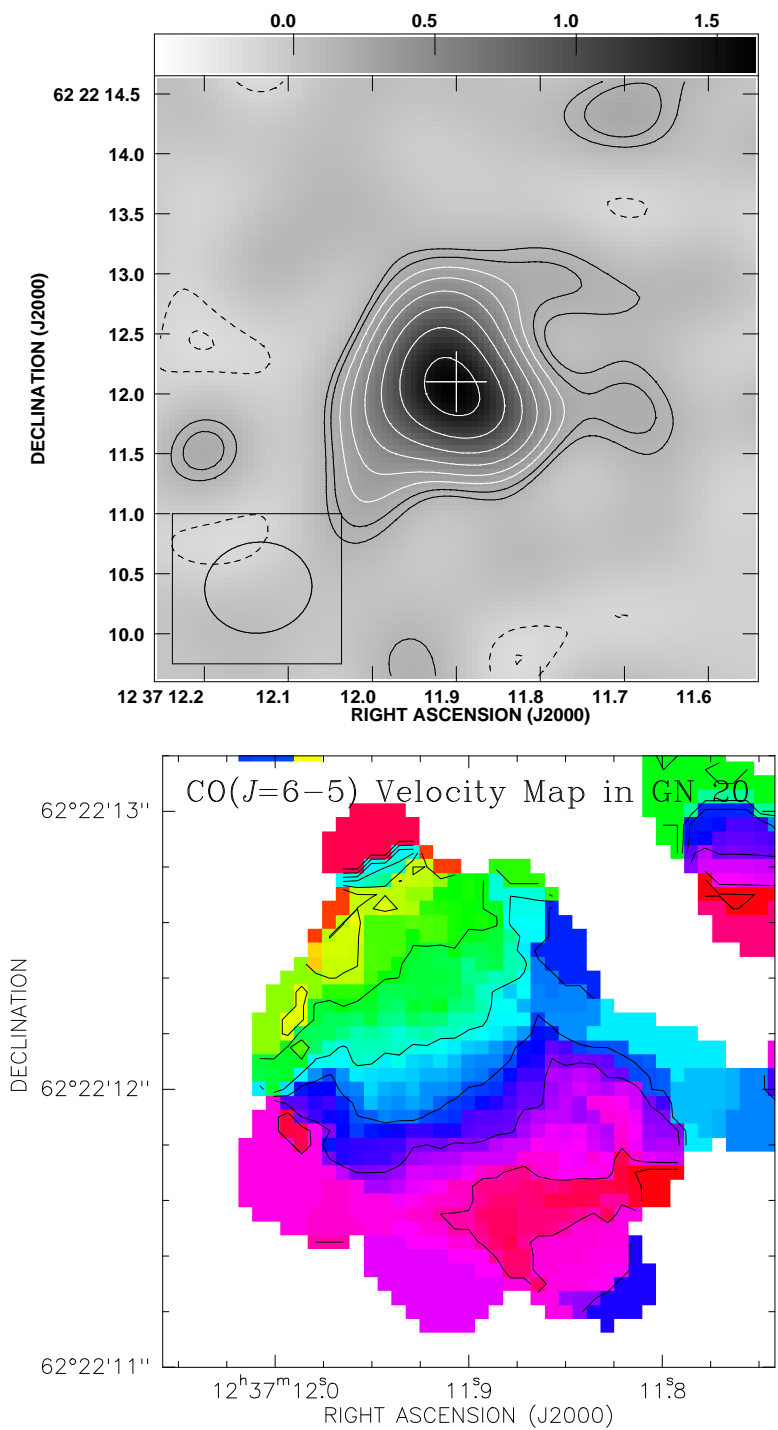,width=4in,angle=-90}
\caption{
CO 6-5 images of GN20 from the Plateau de Bure.  {\bf Upper:} The
total emission integrated over 800 km s$^{-1}$ at $0.89'' 
\times 0.76''$ resolution (NA weighting).  Contour levels are a 
geometric progression in the square root two starting at 
0.13 mJy beam$^{-1}$. The greyscale flux units are in mJy 
beam$^{-1}$. The rms in the image is 0.1 mJy beam$^{-1}$. {\bf
Lower:} The first moment image at $0.84'' \times 0.67''$
resolution (UN weighting), ie. intensity weighted mean velocity at a
given position. Contour levels are in steps of 100 km s$^{-1}$.
}
\end{figure}

\begin{figure}
\psfig{file=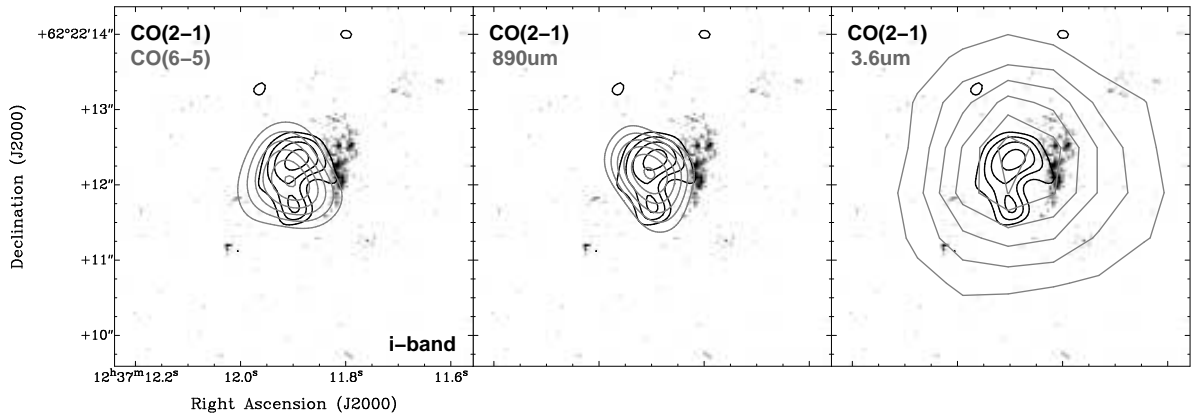,width=7in}
\caption{
{\bf Left:} The dark contours show the CO 2-1 emission from GN20 at
0.45$"$ resolution. The greyscale is the HST+ACS I-band image.  
The light contours show the CO 6-5 emission at $0.8"$ resolution. 
{\bf Middle:}  The light contours show the 850$\mu$m emission
from the SMA at $0.8"$ resolution (Younger et al. 2008). 
{\bf Right:} The contours show the Spitzer
IRAC image at 3.6$\mu$m of GN20 (PSF FWHM=$1.5''$). 
}
\end{figure}

\begin{figure}
\psfig{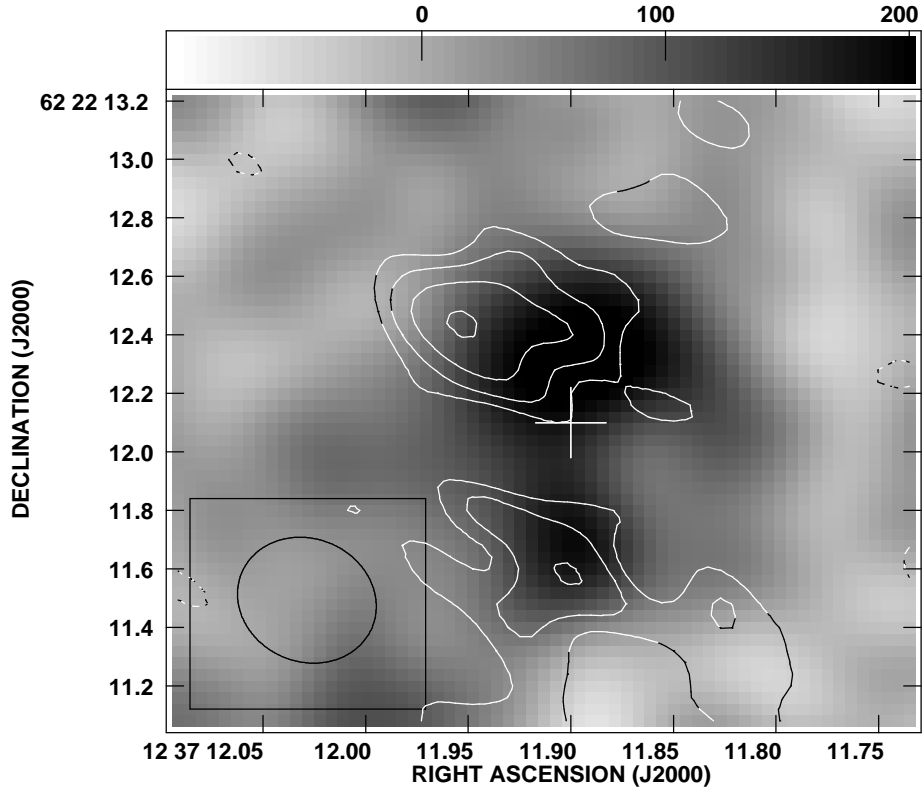}
\caption{
The greyscale is the VLA image of CO 2-1 emission from GN20 at 0.45$''$
resolution. The contours are the VLA+MERLIN 1.4GHz radio image, also
convolved to 0.45$"$ resolution (Casey et al. 2009).  The contour
levels are a geometric progress in square root two, starting at
5.5$\mu$Jy beam$^{-1}$.  Again, the
cross shows the radio peak position of GN20 at 1.7$"$ resolution
(Morrison et al. 2009).
}
\end{figure}

\begin{figure}
\psfig{file=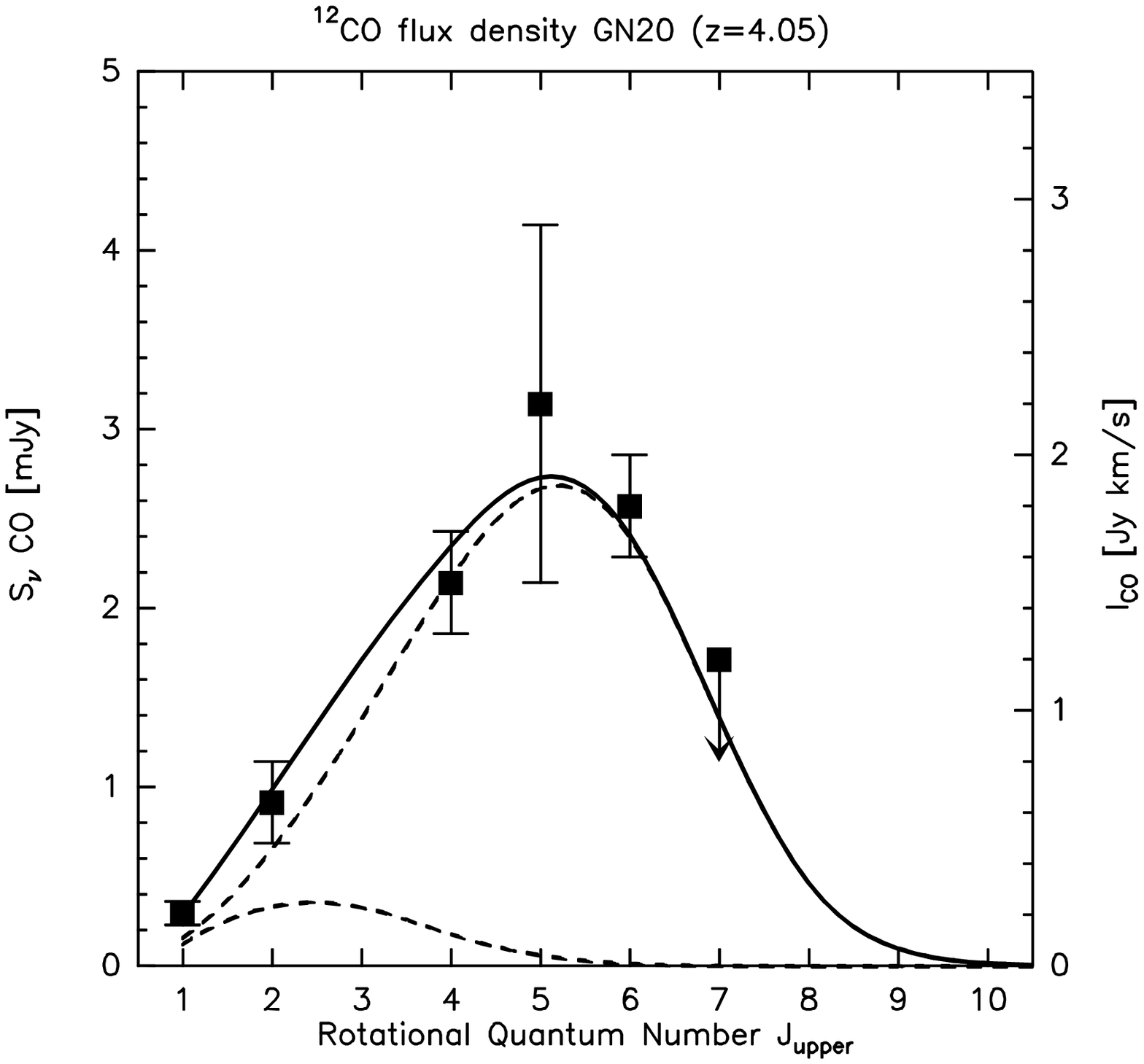,width=3.5in}
\vskip 0.1in
\psfig{file=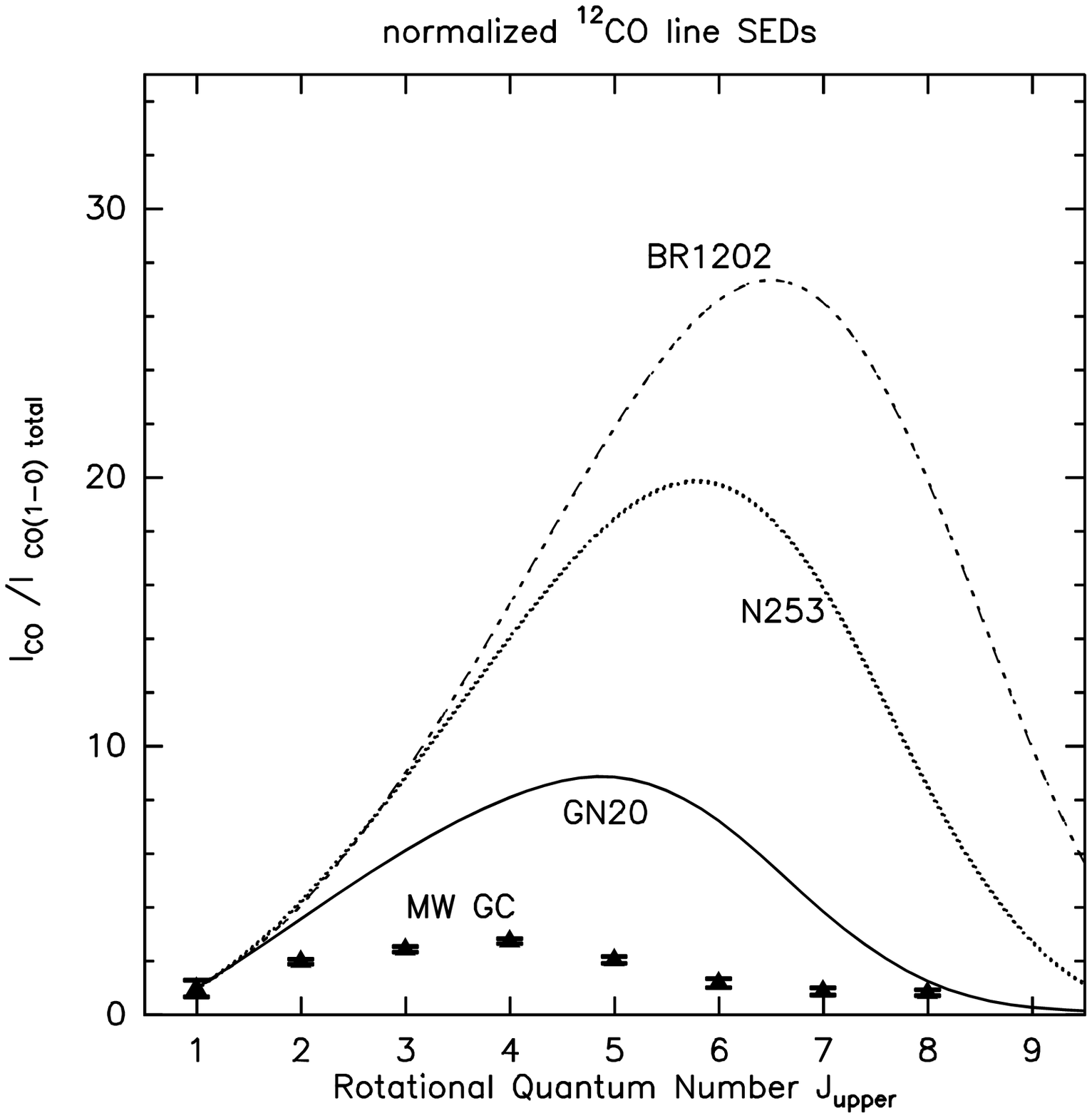,width=3.2in}
\caption{
The CO excitation ladder for GN20. {\bf Upper:} The total CO emission 
from GN20 plus a
double component LVG model, as described in Section 5.3. {\bf Lower:} A
comparison of the GN20 CO excitation with the starburst regions of NGC
253, the FIR hyper-luminous ($L_{FIR} \sim 10^{13}$ L$_\odot$) host
galaxy of the $z = 4.7$ quasar BR1202-0725, and the Milky Way inner
disk.
}
\end{figure}

\end{document}